\documentclass[a4paper,10pt,twoside]{cpc-hepnp}

\usepackage{multicol}
\usepackage{graphicx}
\usepackage{booktabs}
\usepackage{amssymb,bm,mathrsfs,bbm,amscd}
\usepackage[tbtags]{amsmath}
\usepackage{lastpage}
\usepackage{color} 
\usepackage{subfig}
\usepackage{CJK}

\begin{document}
\begin{CJK*}{GBK}{song}

\fancyhead[c]{\small Chinese Physics C~~~Vol. XX, No. XX (2013) 010201} 
\fancyfoot[C]{\small 010201-\thepage}
\footnotetext[0]{Received xx August 2013}

\title{Waveform Digitizing for LaBr$_3$/NaI Phoswich Detector\thanks{Supported by NSFC under grants 10978001 and 11222327.}}
\author{SHE Rui$^{1,2}$,
\quad JI Jian-Feng$^{1,2;1)}$\email{jijianf@tsinghua.edu.cn},
\quad FENG Hua$^{1,2}$, 
\quad LI Hong$^{1,2}$, 
\quad HAN Dong$^{1,2}$
}
\maketitle

\address{%
$^1$ Department of Engineering Physics, Tsinghua University, Beijing 100084, China\\
$^2$ Key Laboratory of Particle \& Radiation Imaging (Tsinghua University), Ministry of Education, China
}

\begin{abstract}
The detection efficiency of phoswich detector starts to decrease when Compton scattering becomes significant. Events with energy deposit in both scintillators, if not rejected, are not useful for spectral analysis as the full energy of the incident photon cannot be reconstructed with conventional readout.  We show that once the system response is carefully calibrated, the full energy of those double deposit events can be reconstructed using a waveform digitizer as the readout. Our experiment suggests that the efficiency of photopeak at 662 keV can be increased by a factor of 2 given our LaBr$_3$/NaI phoswich detector. 
\end{abstract}

\begin{keyword}
X-ray detector, phoswich, LaBr$_3$, waveform digitizing
\end{keyword}

\begin{pacs}
95.55.Ka
\end{pacs}

\footnotetext[0]{\hspace*{-3mm}\raisebox{0.3ex}{$\scriptstyle\copyright$}2013
Chinese Physical Society and the Institute of High Energy Physics
of the Chinese Academy of Sciences and the Institute
of Modern Physics of the Chinese Academy of Sciences and IOP Publishing Ltd}%

\begin{multicols}{2}

\section{Introduction}
Phoswich Detectors, a combination of two scintillation crystals with distinct light decay times, have been widely used in hard X-ray astronomy, in particular where large detection area is needed.  In the past, a combination of NaI(Tl) and CsI(Na), with a decay time of 630 ns and 250 ns respectively, is often used as a phoswich. For example, the Phoswich Detection System (PDS) onboard BeppoSAX \citep{Frontera1997}, which is sensitive in the energy range of 15 to 300 keV, has resulted in fruitful scientific outcome.  Recently, LaBr$_3$ doped with Ce was found to be a new scintillation material with better energy resolution and much shorter decay time (16 ns) than that of NaI \citep{Shah2003}.  Several studies suggest that a phoswich detector in combination with LaBr$_3$ and NaI has great potentials in X-ray astronomy \citep{Mazumdar2010,Manchanda2011,Li2012}, especially for time domain astronomy and transient detection. 

In phoswiches, the top crystal acts as the sensitive volume,  while the bottom crystal is used as an active veto to shield background. The sensitive layer should be much thinner than the veto layer in order for an effective background rejection.  For X-rays with energy above $\sim$200--300 keV, Compton scattering becomes significant in detectors, leading to a low detection efficiency in phoswich due to energy deposit in the bottom layer by Compton events. Even if not rejected, events with energy deposit in both layers cannot be used for spectroscopy due to different decay times and light yields. This makes the phoswich detectors difficult to detect X-rays above 300 keV, and may miss interesting sciences in this energy range, for examples,  the Galactic electron-positron annihilation at 511 keV \citep{Teegarden2005}, or gamma-ray bursts with high peak energies \citep{Mallozzi1995}. 

In a previous study \citep{Li2012}, we have demonstrated the technical readiness of a large area ($\Phi$101.6 mm)  LaBr$_3$:Ce/NaI(Tl) phoswich detector and its potential capability for hard X-ray astronomy . This phoswich detector shows good low energy response (below 5 keV),  fine energy resolution ($\sim$10\% at 60 keV), and excellent spatial unification (less than 1\% in pulse amplitude across the sensitive surface).  In the pulse width spectrum, the two peaks caused by energy deposit in either LaBr$_3$:Ce or NaI(Tl) are distinctly separated as per their large decay time ratio (250:16).

Here in this work, we show that if a waveform digitizer is used as the readout, Compton events with energy deposit in both layers can be recognized and the full energy of incident photons can be reconstructed by fitting the pulse waveform. This can improve the high energy response of phoswich detectors.

\section{Instrument Setup}

The phoswich detector consists of cylindrical LaBr$_3$:Ce and NaI(Tl) scintillators with a diameter of 101.6 mm. The top LaBr$_3$:Ce layer is 6mm-thick and the bottom NaI(Tl) layer is 40mm-thick. The entrance window is 0.22mm-thick beryllium. A photomultiplier tube (PMT; HAMAMATSU R877) is coupled to the detector below NaI(Tl). With simulations using GEANT4,  the detection efficiency versus photon energy is shown in Fig.~\ref{fig:eff}, where the solid curve corresponds to efficiency of detection and the dashed curve is for efficiency of full energy deposit in LaBr$_3$. At 662 keV, the detection efficiency is 22\% while the full energy efficiency is only 5\%. 

% 1. remove legend
% 2. use smooth curves with enough details
\begin{center}
\includegraphics[width=\columnwidth]{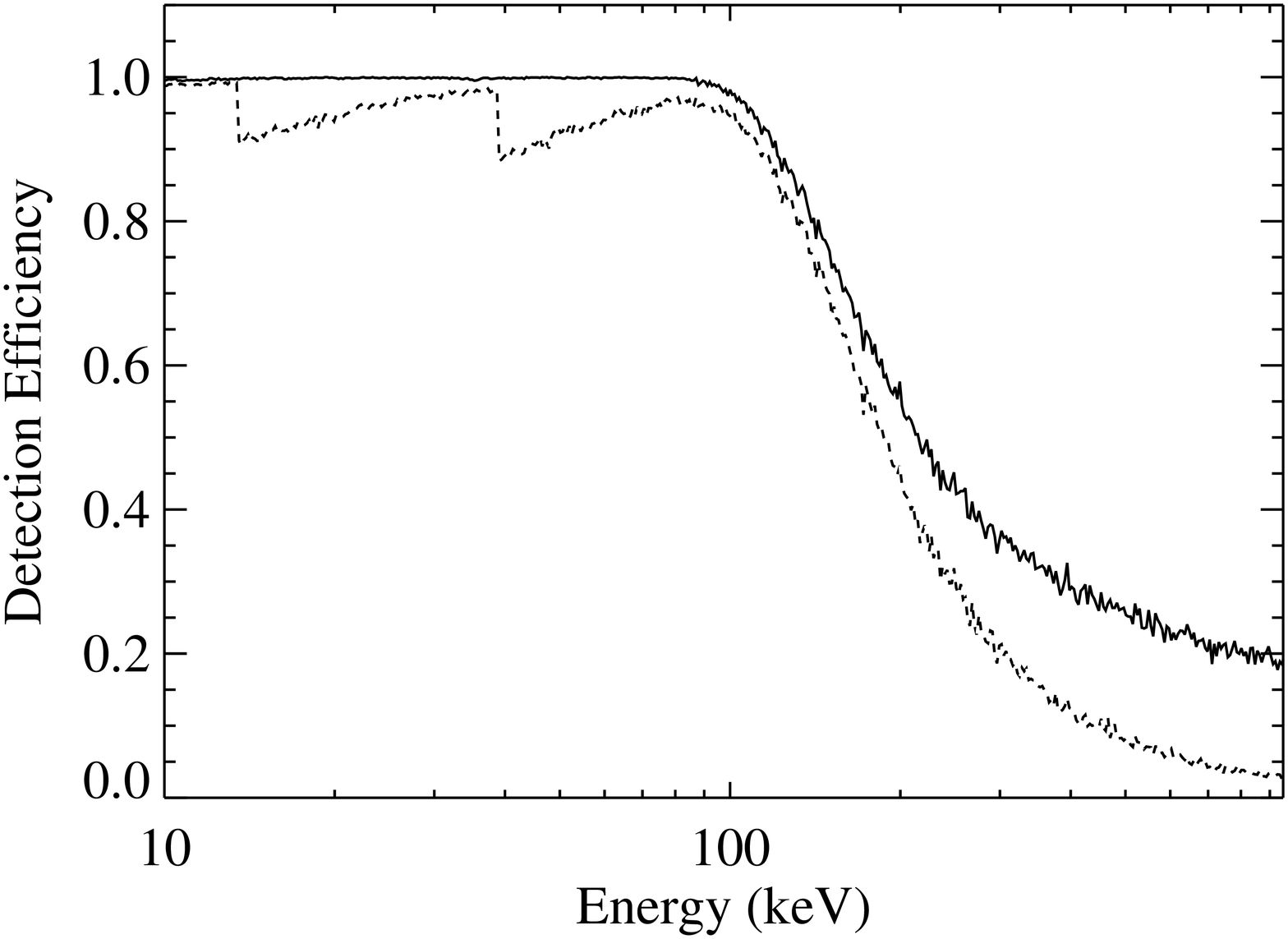}
\figcaption{Simulated quantum efficiency of the phoswich detector as a function of energy.  The solid curve indicates the efficiency of detection, i.e.\ the probability of any amount of energy deposited in LaBr$_3$, while the dashed curve indicates the efficiency of full energy deposit in LaBr$_3$.\label{fig:eff}} 
\end{center}

The PMT anode is connected with a preamplifier with a discharge timescale of roughly 10~ns, whose output is fed into a digitizer (CAEN V1742, 12-bit) sampling at a rate of 1GS/s with a buffer size of 1024. Thus, every event recorded by the digitizer has a length of 1024~ns in time. The PMT was operated at a negative high voltage of 1080~V.  All the experiments were conducted at room temperature.

\section{System Response and Calibration}

The light output in scintillation crystal is expected to be a exponential curve, 
\begin{equation}
  I = I_0~e^{-\frac{t}{\tau}},
\end{equation}
where $\tau$ is the decay time of the scintillator and $I_0$ is the light yield per unit time at the very beginning. Such a signal will go through the PMT and the preamplifier and convolve with their responses. We define their overall response as the system response $R_{\rm s}$. The system response can be measured by illuminating the PMT cathode with a pulsed light source whose pulse width is much smaller than the response time.   A nano-second ultraviolet (UV) LED pulser (HORIBA NanoLED-265) is used to generate light pulses, with a peak wavelength of 268 nm and a full width at half maximum (FWHM) less than 1.2~ns. The LED light is injected to the central position on the PMT cathode and the system response $R_{\rm s}$ is measured by the digitizer, see Fig.~\ref{fig:response}. The response shows a variability of less than 3.5\% in FWHM with the change of PMT high voltage, illumination location and the light luminosity, which is negligible for the analysis.

\begin{center}
\includegraphics[width=\linewidth]{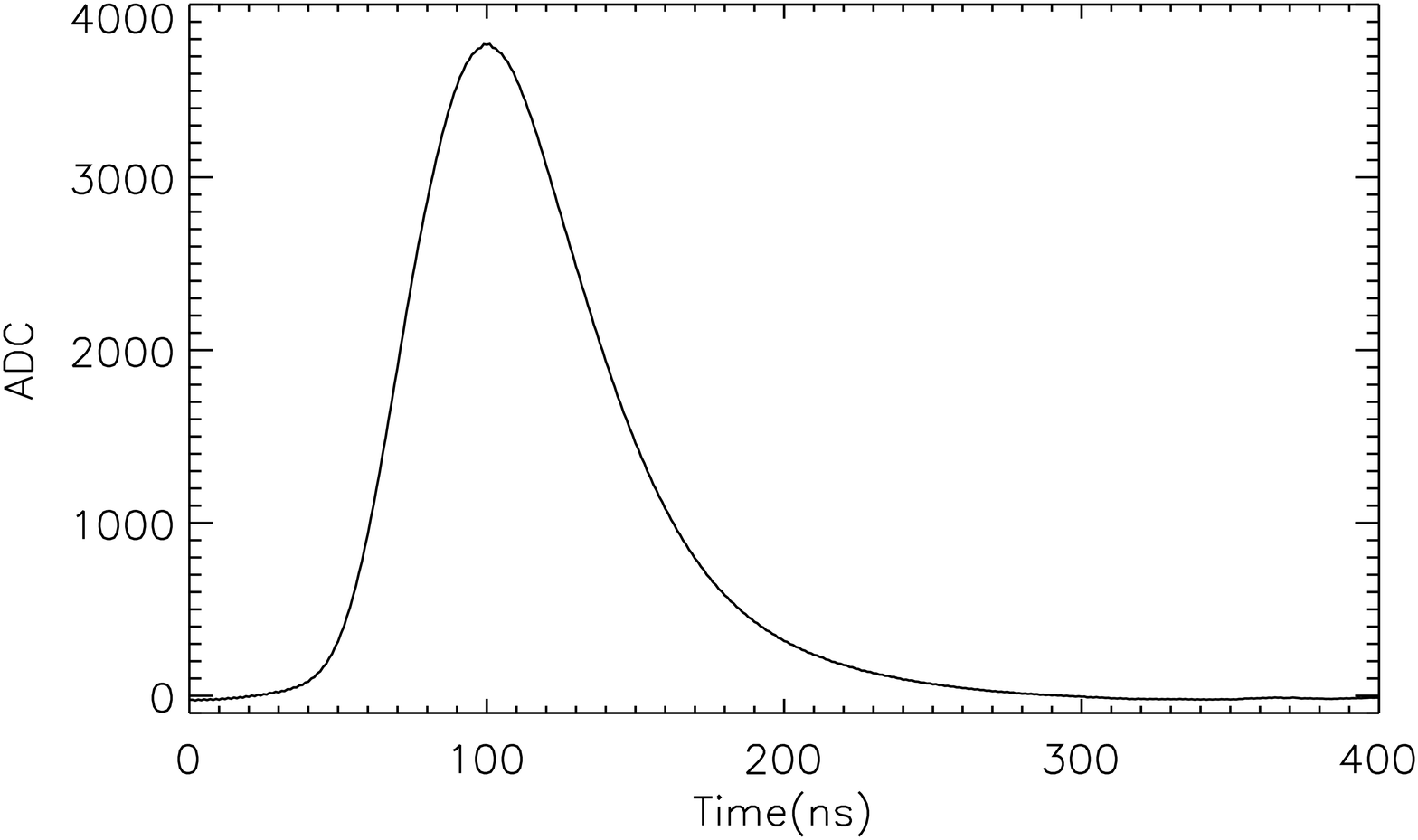}
\figcaption{Measured system response $R_{\rm s}$ using a nano-second UV LED.   \label{fig:response}}
\end{center}

The output signal from the preamplifier for an X-ray event, $S_{\rm o}$ is therefore a convolution of the light decay with the system response,
\begin{equation}
  S_{\rm o}=I \ast R_{\rm s} \; .
\end{equation}

\section{Measurements and Results}

A radioactive source $^{137}$Cs is used to produce 662~keV gamma-rays. A measurement of FWHM versus amplitude for signals from the preamplifier is shown in Fig.~\ref{fig:diagram}. Each event is a point on the left panel. The distribution of pulse width is shown on right, with two peaks corresponding to energy deposit in LaBr$_3$ and NaI, respectively ($\sim$85 ns vs.\ $\sim$290 ns).  There are some events with FWHM in between the two peaks of LaBr$_3$ and NaI, lying on a curve on the FWHM versus peak diagram, which are Compton events with energy deposit in both layers, 

\begin{center}
\includegraphics[width=\columnwidth]{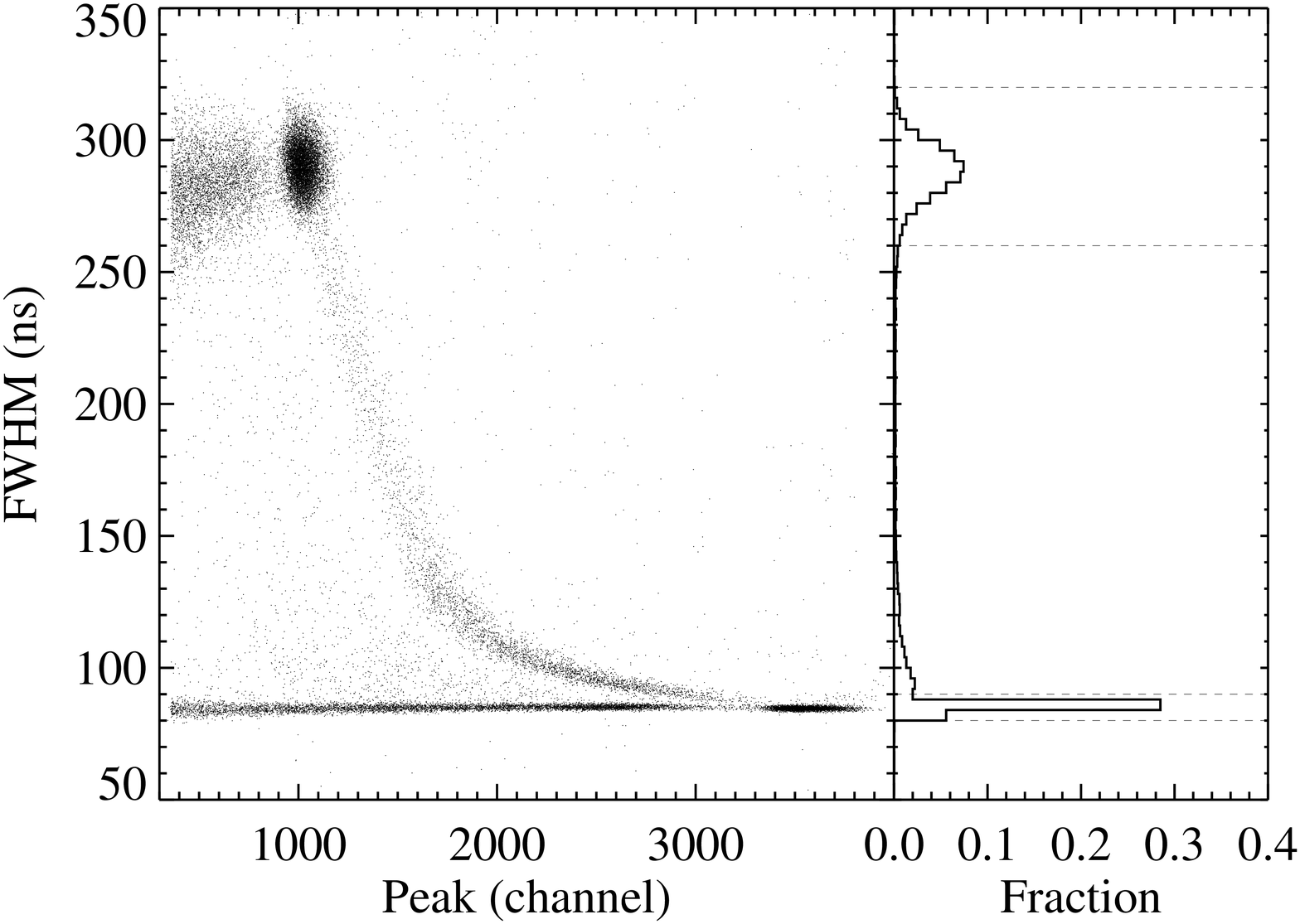}
\figcaption{FWHM versus amplitude (left) for measured events with a $^{137}$Cs at 662~keV. distributions. The right panel shows the distribution of FWHM and the two peaks are events with energy deposit in LaBr$_3$ and NaI, respectively.  The dashed lines define two regions (80--90 ns and 260--320 ns, respectively) for events selection  \label{fig:diagram}}
\end{center}

The energy spectra for pure LaBr$_3$ or NaI events  are shown in Fig.~\ref{fig:spec}. The LaBr$_3$ events are selected for those with FWHM in the range of 80--90 ns, and the NaI events are those in 260--320 ns. The channel-energy calibration is done using radioactive sources $^{60}$Co and $^{241}$Am. The peak in Fig.~\ref{fig:spec} is the full energy deposit from the 662 keV gamma-rays and the broad continuum on its left is from Compton events with partial energy deposit. 

\begin{center}
\includegraphics[width=\columnwidth]{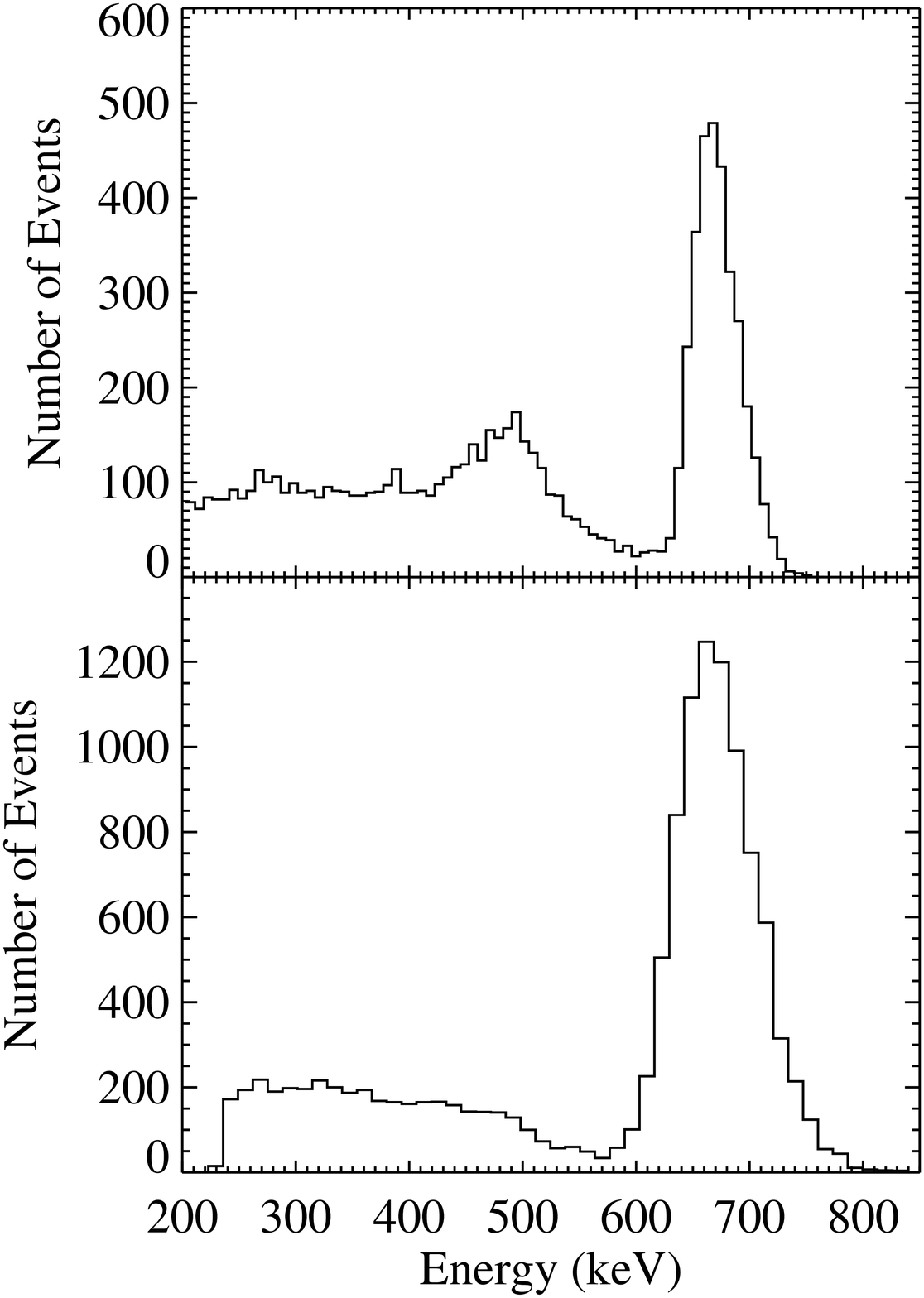}
\figcaption{Energy spectra of $^{137}$Cs for pure LaBr$_3$:Ce (top) or NaI(Tl) (bottom) events.\label{fig:spec}}
\end{center}

For events with FWHM between 90 ns and 260 ns, which have energy deposit in both layers, their signal waveforms are fitted using the following function
\begin{equation}
  I = I_{1,0}~e^{-\frac{t}{\tau_1}} + I_{2,0}~e^{-\frac{t}{\tau_2}},
\end{equation}
where the subscription 1 and 2 denotes LaBr$_3$ or NaI, respectively. The decay times are fixed at their theoretical values, $\tau_1 = 16$~ns and $\tau_2 = 250$~ns. There are only two independent parameters in the fit, $I_{1,0}$ and $I_{2,0}$, corresponding to the energy deposit in each crystal. A typical fitting result is shown in Fig.~\ref{fig:fit},  where the two components are clearly disentangled.  The total energy of the events can thus be reconstructed by adding up their energies in each crystals. 

\begin{center}
\includegraphics[width=\columnwidth]{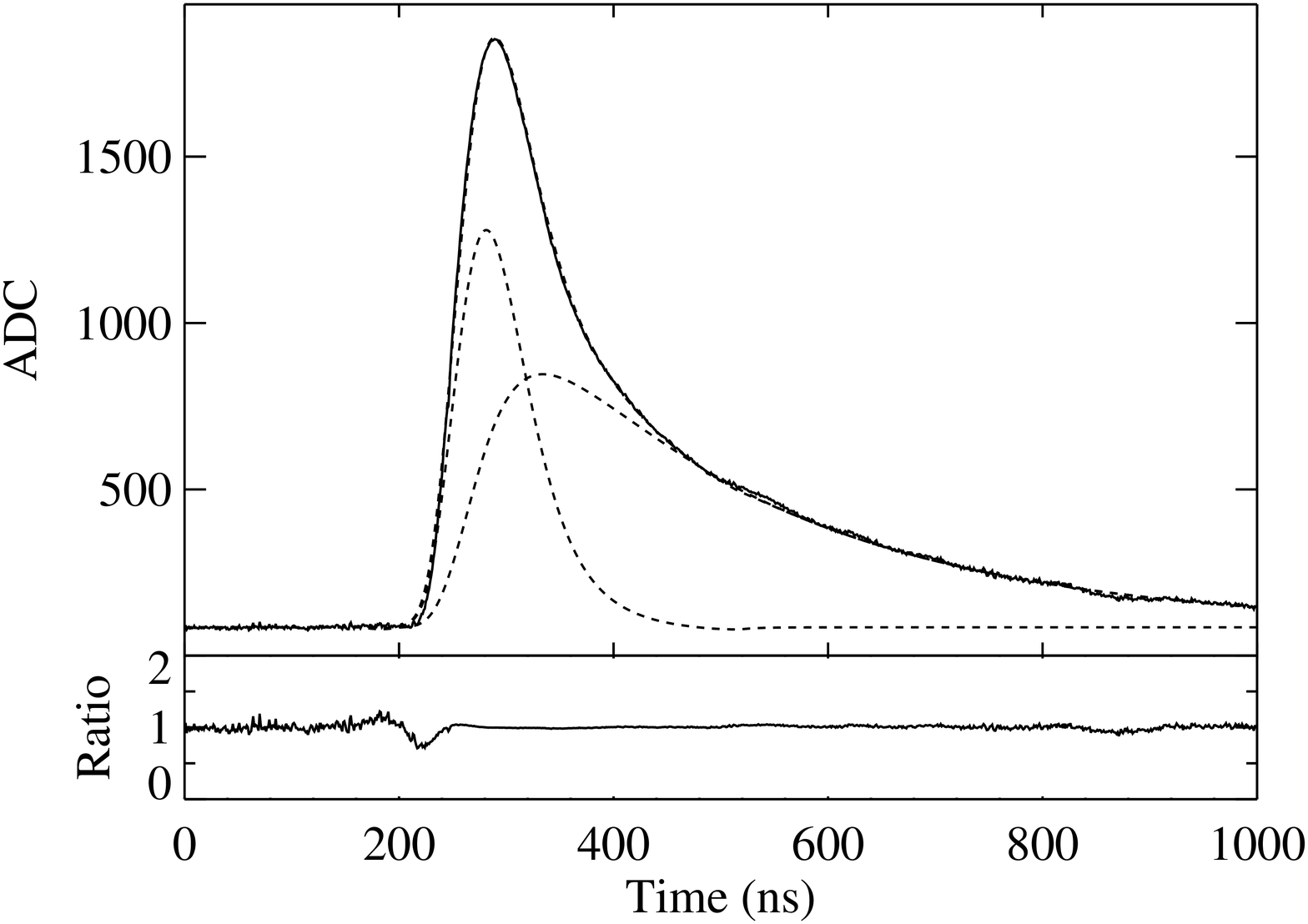}
\figcaption{A typical fitting result of an event with energy deposit in both crystals. The measured and model (total, LaBr$_3$, and NaI) waveforms are shown on top and the data to model ratio is shown on bottom. \label{fig:fit}}
\end{center}

The reconstructed energy spectra for those ``double deposit events'' are shown in Fig.~\ref{fig:sumspec}.  The dashed curve is for pure LaBr$_3$ events and the solid curve is the spectrum after adding up double deposit events.  As one can see, the photopeak at 662~keV is higher but the energy resolution is slightly degraded due to worse resolution of NaI than LaBr$_3$. By including the double deposit events, the efficiency for full energy detection is increased by a factor of 1.9, consistent with estimate using GEANT4 simulation.

\begin{center}
\includegraphics[width=\columnwidth]{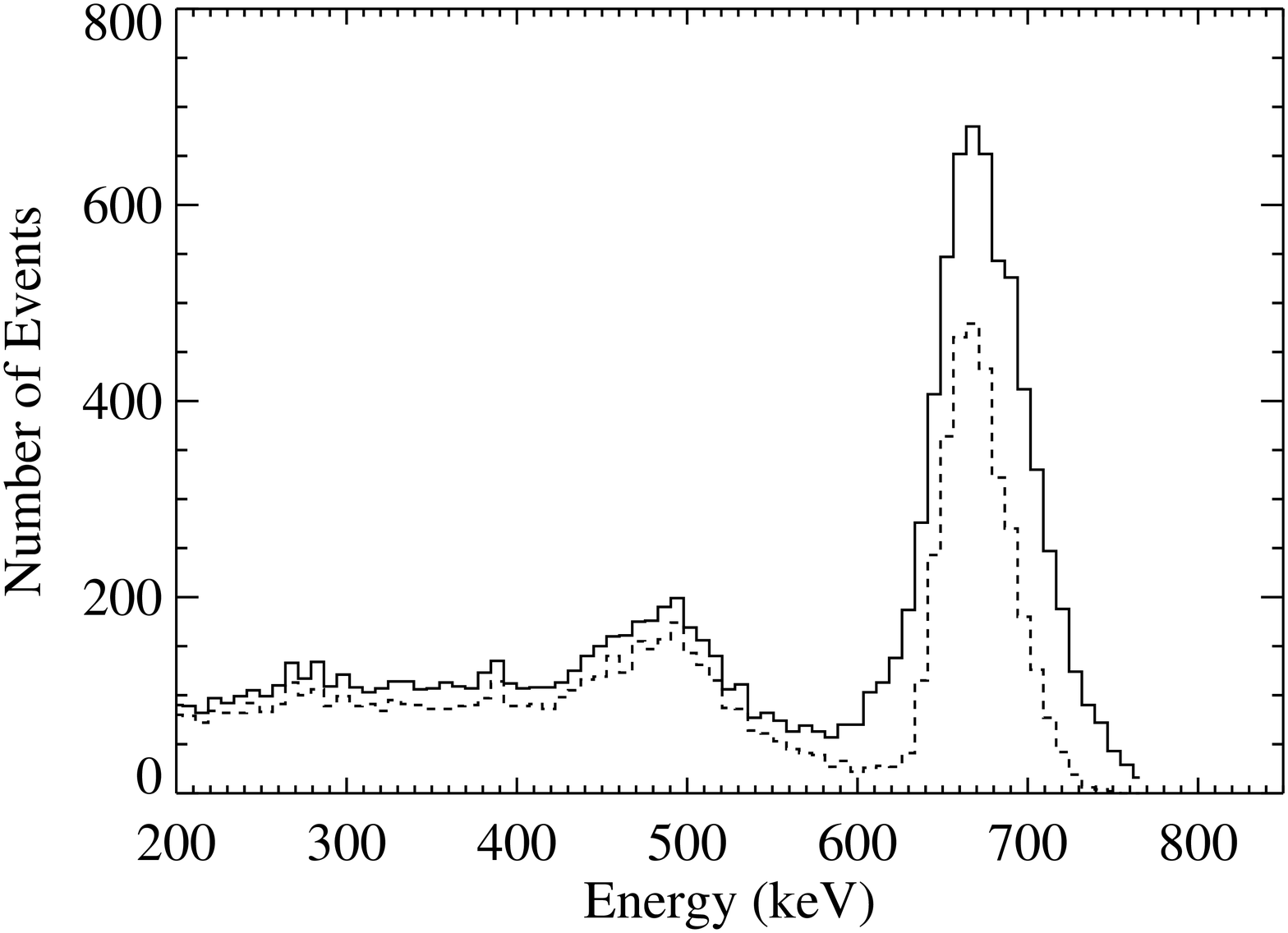}
\figcaption{Energy spectrum of $^{137}$Cs consisting of pure LaBr$_3$ events (dashed) and by adding up double deposit events (solid). \label{fig:sumspec}}
\end{center}

\section{Conclusion}

Here in this work we demonstrate that the Compton events with energy deposit in both LaBr$_3$:Ce and NaI(Tl) can be recognized and the full energy of the incident photon can be reconstructed using a digitizer. This improvement in readout can simply increase the efficiency of full energy detection by a factor of 2, and may broaden the scientific return of phoswich detectors in X-ray and gamma-ray astronomy. 
  
\end{multicols}

\vspace{-1mm}
\centerline{\rule{80mm}{0.1pt}}
\vspace{2mm}

\begin{multicols}{2}

\end{multicols}

\clearpage
\end{CJK*}
\end{document}